\documentclass[fleqn,twoside]{article}
\usepackage{espcrc2}

\usepackage{graphicx}
\usepackage[figuresright]{rotating}

\newcommand{\AmS}{{\protect\the\textfont2
  A\kern-.1667em\lower.5ex\hbox{M}\kern-.125emS}}

\title{Neutrinoless Double Beta Decay: Present and Future}

\author{Oliviero Cremonesi\thanks{e-mail: oliviero.cremonesi@mib.infn.it}
\vspace{1pc}
	\\
	Istituto Nazionale di Fisica Nucleare, \\
        Sezione di Milano, \\
	Piazza della Scienza, 3\\ 
        I-20126 Milano, Italy}
       
\begin{document}

\def\mee{$\langle m_{ee} \rangle$~}
\def\mnu{$\langle m_{\nu} \rangle$~}
\def\mmod{$\| \langle m_{ee} \rangle \|$}
\def\mb{$\langle m_{\beta} \rangle$~}
\def\BBz{$\beta\beta(0\nu)$~}
\def\BBm{$\beta\beta(0\nu,\chi)$~}
\def\BBd{$\beta\beta(2\nu)$~}
\def\BB{$\beta\beta$~}
\def\Mz{$|M_{0\nu}|$~}
\def\Md{$|M_{2\nu}|$~}
\def\Tz{$T^{0\nu}_{1/2}$~}
\def\Td{$T^{2\nu}_{1/2}$~}
\def\Tm{$T^{0\nu\,\chi}_{1/2}$~}
\def\ca{$\sim$}
\def\dca{$\approx$}
\def\dot{$\cdot$}
\def\teod{TeO$_2$~}

\begin{abstract}
Present status, and future plans for Double Beta Decay searches are reviewed. 
Given the recent observations of neutrino oscillations, a possibility to observe \BBz at a neutrino mass scale suggested by present experimental results ($m_{\nu} \approx$10-50 meV) could actually exist. The achievement of the required experimental sensitivity is a real challenge faced by a series of new proposed projects. Plans to achieve such a result are described.
\vspace{1pc}
\end{abstract}

\maketitle

\section{Introduction}
The recent results from the neutrino oscillation experiments convincingly show that neutrinos have a finite mass. However, in such experiments only the squares of the neutrino masses differences can be measured, and only a lower limit on the absolute value of the neutrino mass scale has been obtained in this way. Such a result is in turn causing a renewed interest in double beta decay experiments which are expected to reach, in the next future, a sensitivity corresponding to the mass scale indicated by neutrino oscillation experiments. 
\BBz is actually a very important process both from the particle and nuclear physics point of view, representing a unique tool to establish the absolute neutrino mass scale, its nature (Dirac/Majorana) and the values of the Majorana CP phases. It can proceed in fact only if neutrinos are Majorana massive particles. Unfortunately, uncertainties in the transition nuclear matrix elements still affect the interpretation of the experimental results and new efforts to overcome such a problem are strongly required (new theoretical calculations and experimental analyses of different \BBz active isotopes). Complementary informations from beta experiments and astrophysics are of course welcomed.\\
Present, and near future \BBz experiments have reached a sensitivity to span the \mnu region 0.1-1 eV. Different plans to overcome the challenge implied by the achievement of the sensitivity required to reach the 10-50 meV region have been proposed. They will be reviewed in the section devoted to the future projects.
Several review articles covering both the experimental and the theoretical aspects and implications of \BB have been recently issued \cite{ELL02,VER02,SUH98,VIS02,TRE02}. I will refer to them for more details on the subject.

\section{Double Beta Decay}

\begin{figure}[bt]
    \begin{minipage}[c]{0.5\textwidth}%
      \centering\includegraphics[height=3.5 cm]{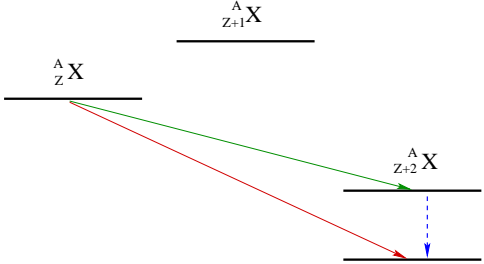}%
    \end{minipage}%
\caption{Simplified scheme of the \BB transitions.}
\label{fig:bbs}
\end{figure}

Double Beta Decay is a rare spontaneous nuclear transition in which the charge of two isobaric nuclei changes by two units with the simultaneous emission of two electrons. The parent nucleus must be less bound than the daughter one, while it is generally required that both be more bound than the intermediate one, in order to avoid the equivalent sequence of two single beta decays (Fig.\ref{fig:bbs}). These conditions are fulfilled in nature for a number of even-even nuclei. The decay can then proceed both to the ground state or to the first excited states of the daughter nucleus. 
Nuclear transitions accompanied by positron emission or electron capture processes are also possible. They are however characterized by poorer  experimental sensitivities and will not be discussed in the following.
Several \BB modes are possible. The most popular are the 2$\nu$ mode 
\begin{equation}
^A_ZX \to ^A_{Z+2}X + 2e^- + 2\overline{\nu}
\end{equation}
which conserves the lepton number and it is allowed in the framework of the Standard Model (SM) of electro-weak interactions, and the 0$\nu$ mode 
\begin{equation}
^A_ZX \to ^A_{Z+2}X + 2e^-
\end{equation}
which violates the lepton number and has been recognized since a long time as a powerful tool to test neutrino properties\cite{FUR39}. A third decay mode (\BBm), in which one or more light neutral bosons $\chi$ (Majorons, whose existence is postulated by various SM extensions) are also emitted
\begin{equation}
^A_ZX \to ^A_{Z+2}X + 2e^- + N\chi
\end{equation}
is often considered.
In all envisaged modes, \BB is a second order weak semileptonic transition, hence characterized by very long lifetimes. Besides the exchange of light or heavy Majorana neutrinos, \BBz can be mediated by the exchange of a variety of unconventional particles (e.g. SUSY partners). Its amplitude then depends on the mass and coupling constants of these virtual particles and \BBz results can be used to constrain model parameters. Important constraints on Left-Right Symmetric and Supersymmetric models have thus been obtained using currently available \BBz results.\\
Independent of the actual mechanism mediating the decay, \BBz observation would necessarily imply that neutrinos are Majorana massive particles. A lower limit on neutrino mass eigenstates could be then obtained. Lacking however any evidence for \BBz experimental upper limits on the decay lifetimes can only be interpreted as independent limits on each of the possible contributions to the decay amplitude.
Disregarding more unconventional contributions (SUSY or left-right symmetric models), the \BBz rate is usually expressed as
\begin{equation}
[ T_{1/2}^{0\nu}]^{-1} = 
G^{0\nu}|M^{0\nu}|^2{\langle m_{\nu} \rangle^2}
\label{eq:rate0}
\end{equation}
where $G^{0\nu}$ is the (exactly calculable) phase space integral, $|M^{0\nu}|^2$ is the nuclear matrix element and \mnu ({em effective neutrino mass}) is the neutrino relevant parameter measured in \BBz
\begin{equation}
\label{eq:effterms}
\langle m_{\nu} \rangle = \sum_k \phi_k m_k U_{e,k}^2 
\end{equation}
Here, $U$ is the left-handed unitary neutrino mixing matrix relating physical {\em weak eigenstates} to the mass eigenstates ($\nu_L ~=~ U N^L$), while $\phi_k$ are the intrinsic CP parities of the neutrinos. Summation is over light neutrinos only. The presence of the phases $\phi_k$ implies that cancellations are possible. Such cancellations are complete for a Dirac neutrino since it is equivalent to two degenerate Majorana neutrinos with opposite CP phases. This stresses once more the fact that \BBz can occur only through the exchange of Majorana neutrinos.\\
From a Particle Physics point of view, \BBz represents a unique tool to measure the neutrino Majorana phases and to assess the absolute scale of the neutrino masses. Predictions on \mnu based on the most recent neutrino oscillation results have been derived by various authors~\cite{PET02,SMI02}. The most striking aspect of such predictions is that, for the first time in the history of \BB searches, a definite goal exists: an experimental sensitivity in the range \mnu \ca10-50 meV could definitely rule out inverse and quasi-degenerate hierarchies thus assessing a direct neutrino mass hierarchy~\cite{PET02}.
\BBz observation at larger \mnu scales would be equally important but its occurrence is based on more optimistic assumptions (e.g. a definite mass hierarchy).\\
As it is apparent from eq.~\ref{eq:rate0} the derivation of the crucial parameter \mnu from the experimental results on \BBz lifetime requires a precise knowledge of the transition Nuclear Matrix Elements (NME). Unfortunately this is not an easy job and a definite knowledge of NME values and uncertainties is still lacking in spite of the large attention attracted by this area of research. Many, often conflicting evaluations are available in the literature and it is unfortunately not easy to judge their correctness or accuracy. 
Outstanding progress has been achieved over the last years mainly due to the application of the QRPA method and its extensions. Renewed interest in Shell Model calculations has been on the other hand boosted by the fast development of computer technologies. Alternative approaches (e.g. OEM) have also been pursued.
Comparison with experimental \BBd rates has often been suggested as a possible way out (direct test of the calculation method). The evaluation methods for the two decay modes show however relevant differences (e.g. the neutrino propagator) and the effectiveness of such a comparison is still controversial. 
A popular even if doubtful attitude consists in considering the spread of the different evaluations as an estimate of their uncertainties. In such a way one obtains a spread of about one order of magnitude in the calculated half lifetimes (Tab. \ref{tab:NME}), corresponding to a factor of \ca~3 in \mnu. It is clear that a big improvement in the calculation of NME or at least in the estimate of their uncertainties would be welcomed. New calculation methods should be pursued while insisting on the comparison with dedicated measurements coming from  various areas of nuclear Physics~\cite{EJI01}. On the other hand, an experimental effort to investigate as many as possible \BB emitters should be addressed.
\begin{table}[htb]
\caption{Theoretically evaluated \BBz\ half-lives (units of $10^{28}$ years for
\mnu\ = 10 meV).}
\label{tab:NME}
\begin{tabular}{@{}lcccccc}
\hline
Isotope & \cite{HAX84} & \cite{CAU99} & \cite{ENG88} & \cite{STA90}
& \cite{FAE98} & \cite{PAN96} \\
\hline
$^{48}$Ca & 3.18    & 8.83   &   -   &   -  &    -  & 2.5\\
$^{76}$Ge & 1.7 & 17.7 & 14.0 & 2.33 & 3.2 & 3.6\\
$^{82}$Se & 0.58 & 2.4 & 5.6 & 0.6 & 0.8 & 1.5\\
$^{100}$Mo & - & - & 1.0 &  1.28 & 0.3  & 3.9\\
$^{116}$Cd &  -   &   -   &  -   &  0.48   & 0.78     & 4.7\\
$^{130}$Te & 0.15 & 5.8 & 0.7 & 0.5 & 0.9 & 0.85\\
$^{136}$Xe & - & 12.1 & 3.3 & 2.2 & 5.3 & 1.8\\
$^{150}$Nd &  -   &  -    &   -   & 0.025    &  0.05    & -\\
$^{160}$Gd  & -    &  -    &   -   & 0.85    &	-   &  -\\
\hline
\end{tabular}
\end{table}

\section{Experimental approaches}
Two main general approaches have been followed so far to investigate \BB:
\begin{description}
\item[i)] indirect or inclusive methods;
\item[ii)] direct or counter methods.
\end{description}
Inclusive methods are based on the measurement of anomalous concentrations of the daughter nuclei in properly selected samples, characterized by very long accumulation times. They include Geochemical and Radiochemical methods and, being completely insensitive to different \BB modes, can only give indirect evaluations of the \BBz and \BBd lifetimes. They have played a crucial role in \BB searches especially in the past.\\
\begin{figure}[bt]
    \begin{minipage}[c]{0.5\textwidth}%
      \centering\includegraphics[height=6cm]{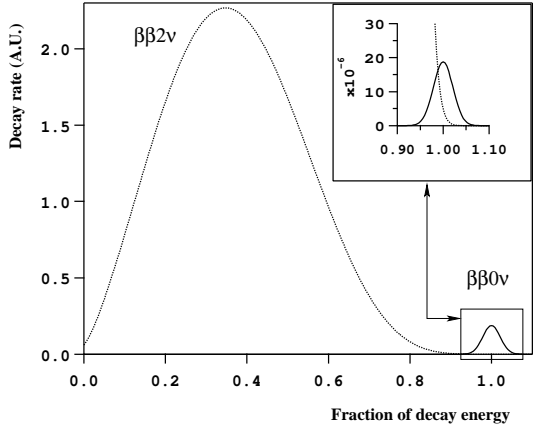}%
    \end{minipage}%
\caption{Electron sum energy spectra for \BBd and \BBz. The relative intensity of \BBd was increased in the inset to stress its contribution to \BBz background.}
\label{fig:espc}
\end{figure}
Counter methods are based instead on the direct observation of the two electrons emitted in the decay. Different features of the event (energies, momenta, topology, etc) are registered according to the different capabilities of the employed detectors. They are further classified in {\em passive} (when the observed electrons are originated in an external sample) and {\em active source} experiments (when the source  of \BB's serves also as detector). The various \BB modes are separated by the differences in their electron sum energy spectra (Fig. \ref{fig:espc}). Because a sharp line at the transition energy is expected for \BBz electron sum energy, direct counting experiments with very good energy resolution are presently attracting the attention of most researchers.
Experimental evidence for several \BBd decays as well as improved lower limits on the lifetimes of many \BBz emitters (Tab. \ref{tab:BBres}) have been provided using the measured two-electron sum energy spectra, the single electron energy distributions and, in some cases, the tracking of the observed particles. 
Various different conventional counters have been used in \BB direct searches: solid state devices (Germanium spectrometers and Silicon detector stacks), gas counters (time projection chambers, ionization and multiwire drift  chambers) and scintillators (crystal scintillators and stacks of plastic scintillators). New techniques based on the use low temperature {\em true} calorimeters have been on the other hand proposed and developed in order to improve the experimental sensitivity and enlarge the choice of suitable candidates for \BB searches investigable with an {\em active source} approach.
A common feature of all \BB experiments has been the constant fight against backgrounds caused mainly by environmental radioactivity, cosmic radiation and residual radioactive contaminations of the detector setup elements. The further suppression of such backgrounds will be the actual challenge for future projects whose main goal will be to maximize \BBz rate while minimizing background contributions.\\
In order to compare the performance of present and future \BB experiments let us introduce an experimental {\em sensitivity} or detector {\em factor of merit}, defined as the process half-life corresponding to the maximum signal (N$_B$) that could be hidden by the background fluctuations, at a given statistical C.L. By considering a constant background level (B) which scales linearly with time (T) and detector mass (M), the expected number of background counts in an energy interval equal to the FWHM energy resolution centered around the transition energy is $N_B=B~\Delta E~T~M$. Thus, at 1$\sigma$ level ($n_B=\sqrt{N_B}$)
\begin{eqnarray}
\label{eq:sensitivity}
F_{0\nu} & = &\tau^{Back.Fluct.}_{1/2}=
\ln 2~N_{\beta\beta}\epsilon\frac{T}{n_B} \nonumber \\
& = &\ln 2\frac{x ~\eta ~ \epsilon ~ N_A}{A} 
\sqrt{ \frac{ M ~ T }{B ~ \Delta E} } ~ (68\% C.L.)
\end{eqnarray}  
where N is the number of \BB decaying nuclei under observation, $\eta$ their isotopic abundance, N$_A$ the Avogadro number, A the compound molecular mass, $x$ the number of \BB atoms per molecule, and $\epsilon$ the detection efficiency.\par
Despite its simplicity, equation (\ref{eq:sensitivity}) has the unique advantage of emphasizing the role of the essential experimental parameters: mass, measuring time, isotopic abundance, background level and detection efficiency.\\
$F_{0\nu}$ can be thought as the inverse of the minimum rate which can be detected  in a period T of measurement. When a {\em zero background level} is reached (i.e. {\underline no counts} recorded  in the relevant energy interval over a statistically significant period of time), the term $n_B$ in eq. \ref{eq:sensitivity} is constant (e.g. 2.3 at 90~\% C.L.) and one gets a factor-of-merit which scales linearly with T and the detector mass:
\begin{equation}
\label{eq:sensitivity0}
F^{0}_{0\nu} = \ln 2~N_{\beta\beta}\frac{T}{2.3} ~ (90\% C.L.) ~.
\end{equation}  
Using eq.\ref{eq:rate0} one can then easily obtain the experimantal sensitivity on \mnu
\begin{eqnarray}
F_{\langle m_{\nu} \rangle} & = & \sqrt{\frac{1}{F^{Exp}_{0\nu}G_{0\nu}|M_{0\nu}|^2}}\\
& = & \left[ \frac{A}{x \eta \epsilon N_A G^{0\nu} |M_{0\nu}|^2} \right]^{1/2}
\left[ \frac{B \Delta E}{M T} \right]^{1/4} \nonumber \\
F^{0}_{\langle m_{\nu} \rangle} & = & 
\left [ \frac{A}{x \eta \epsilon N_A G^{0\nu} |M_{0\nu}|^2} \right]^{1/2}
\times \frac{1}{\sqrt{M T}} 
\label{eq:sensitivitym}
\end{eqnarray}
It is now clear that only future projects characterized by very large masses (possibly isotopically enriched), good energy resolutions and extremely low background levels will have an actual chance to reach the \mnu region around 10 meV, even if the selection of favourable \BB nuclei and the use of special techniques to suppress background (e.g. topological informations) will help in reaching the goal.
In particular, the effectiveness in reaching the estimated background levels will be the actual measure of a given experiment chances. Extreme care will have to be dedicated to all possible background contributions including environmental radioactivity, cosmogenically and artificially induced activity, natural activity of the setup materials and \BBd.

\begin{table*}[htb]
\caption{Best reported results on \BB processes. Limits are at 90\% C.L. except when noted. \BBd results are averaged over different experiments.
The effective neutrino mass limits and ranges are those deduced by the authors ($\langle m_{\nu}\rangle$) or according to Table~\ref{tab:NME} ($\langle m^\dagger_{\nu}\rangle$).}
\label{tab:BBres}
\newcommand{\m}{\hphantom{$-$}}
\newcommand{\cc}[1]{\multicolumn{1}{c}{#1}}
\renewcommand{\tabcolsep}{0.9pc} 
\begin{tabular}{@{}lllll}
\hline
Isotope 	& T$_{1/2}^{2\nu}$ (y)           & T$_{1/2}^{0\nu}$ (y)                       &$\langle m_{\nu}\rangle$ (eV) &  $\langle m^\dagger_{\nu}\rangle$ (eV)\\ 
\hline
$^{48}$Ca          & $(4.2 \pm 1.2)\times 10^{19}$\cite{BRU00}	&
	$>9.5\times 10^{21} (76\%)$\cite{YOU91}                & $<8.3$ & $<16-30$\\
$^{76}$Ge          & $(1.3 \pm 0.1)\times 10^{21}$\cite{HMO01,AAL96} &
	$>1.9\times 10^{25}$\cite{HMO01}                       & $<0.35$ & $<0.3-1$\\
                   & &
	$>1.6\times 10^{25}$ \cite{AAL99,IGE02}                      & $<0.33-1.35$ &\\
$^{82}$Se          & $(9.2 \pm 1.0) \times 10^{19}$\cite{ELL92,ARN98} &
	$>2.7\times 10^{22}(68\%)$ \cite{ELL92}                & $<5$ & $<4.6-14.4$\\
$^{96}$Zr  & $(1.4^{+3.5}_{-0.5})\times 10^{19}$\cite{ARN99,WIE01} & & & \\
$^{100}$Mo         & $(8.0 \pm 0.6)\times 10^{18}$\cite{DAS95,DES97,ASH01}&
	$>5.5\times 10^{22}$\cite{EJI01b}                      & $<2.1$ & $<2.3-8.4$ \\
$^{116}$Cd         & $(3.2 \pm 0.3)\times 10^{19}$\cite{ARN96,DAN00,EJI95}&
	$>7\times 10^{22}$\cite{DAN00}                         & $<2.6$ & $<2.6-8.2$\\
$^{128,130}$Te     &  &
	Geoch. ratio\cite{BER93}  & $<1.1-1.5$ &\\
$^{128}$Te         & $(7.2\pm 0.3)\times 10^{24}$\cite{BER93,CRU93}&	
	$>7.7\times 10^{24}$ \cite{BER93}                      & $<1.1-1.5$ &\\ 
$^{130}$Te         & $(2.7\pm 0.1)\times 10^{21}$\cite{BER93}&
	$>2.08\times 10^{23}$                      & $<0.9-2.0$ & $<0.85-5.3$\\
$^{136}$Xe         & $>8.1\times 10^{20}$\cite{GAV00}&
	$>4.4\times 10^{23}$ \cite{LUE98}                      & $<1.8-5.2$ & $<2-5.2$\\
$^{150}$Nd         & $7.0_{-0.3}^{+11.8} \times$ 10$^{18}$\cite{DES97,ART95}&
	$>1.2\times 10^{21}$\cite{DES97}                       & $<3$ & $<4.6-6.5$\\
$^{238}$U$^{(3)}$      & $(2.0\pm 0.6)\times 10^{21}$\cite{TUR91} & & & \\
\hline
\end{tabular}
\end{table*}

\section{Present and past experiments}
Impressive progress has been obtained during the last years in improving \BBz half-life limits for a number of isotopes and in sistematically cataloging \BBd rates (Tab. \ref{tab:BBres}). Although \BBd results are in some cases inconsistent, the effort to cover as many as possible \BB nuclei thus allowing a diret check for \BBd NME elements is evident. 
\begin{figure}[bt]
    \begin{minipage}[c]{0.5\textwidth}%
      \centering\includegraphics[height=5 cm]{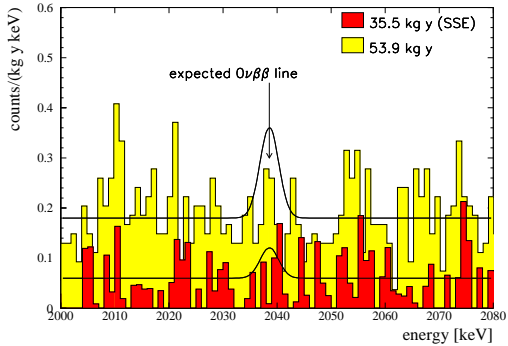}%
    \end{minipage}%
\caption{Heidelberg-Moscow spectra in the \BBz region.}
\label{fig:HMspc}
\end{figure}
Optimal \BBz sensitivities have been reached in a series of experiments based on the calorimetric approach. In particular, the best limit on \BBz comes from the Heidelberg-Moscow (HM) experiment\cite{HMO01} on  $^{76}$Ge (Fig. \ref{fig:HMspc}) even if similar results have been obtained also by  the IGEX experiment\cite{IGE02} (Tab. \ref{tab:BBres}). In both cases a large mass (several kg) of isotopically enriched Germanium diodes (86 \%), is installed deep underground under heavy shields for gamma and neutron environmental radiation. Extremely low background levels are then achieved thanks to a careful selection of the setup materials and  further improved by the use of pulse shape discrimination (PSD) techniques. Both experiments quote similar background levels in the \BBz region of \dca~0.2 (c/keV\dot~kg\dot~y) and \dca~0.06 (c/keV\dot~kg\dot~y) before and after PSD. Taking into account the uncertainties in the NME calculations, such experiments indicate a range of 0.3-1 eV for \mnu. 
As will be discussed later, new ideas to improve such a successful technique characterize many of the proposed future projects.  However, given the NME calculation problem, more \BB emitters than allowed by the use of conventional detectors (e.g. $^{76}$Ge, $^{136}$Xe, $^{48}$Ca) should be investigated using the the calorimetric approach. A solution to this problem, suggested\cite{FIO84} and developed \cite{ALEPE} by the Milano group, is based on the use of low temperature calorimeters ({\em bolometers}). Besides providing very good energy resolutions they can in fact practically eliminate any constraint in the choice of the \BB emitter. Due to their very simple concept (a massive absorber in thermal contact with a suitable thermometer measuring the temperature increase following an energy deposition), they are in fact constrained by the only requirement of finding a compound allowing the growth of a diamagnetic and dielectric crystal. Extremely massive detectors can then be built, by assembling large crystal arrays. Thermal detectors have been pioneered by the Milano group for $^{130}$Te (chosen, because of its favourable nuclear factor-of-merit and large natural isotopic abundance, within a large number of other successfully tested \BB emitters) in a series of constantly increasing mass experiments carried out at {\em Laboratori Nazionali del Gran Sasso} (LNGS), whose last extension is the MIBETA experiment\cite{ALE00}. Consisting of an array of 20 TeO$_2$ crystals totalling a mass of 6.8 kg and operating at a temperature of \ca~12 mK, MIBETA has been characterized by a good energy resolution (8 keV on the average at the \BBz transition energy, 2528 keV) over a long running period (\ca~2 years) and a background level of \ca~0.3 (c/keV\dot~kg\dot~y). The quoted limit of $2.08\times10^{23}$ on the $^{130}$Te \BBz half-life, corresponds to a range of 0.9-2 eV in \mnu which is the best after those indicated by Ge diodes experiments.\\
Half-way with next generation experiments, NEMO III is a passive source detector whose construction has just been completed in the Frejus underground laboratory\cite{NEM02} at a depth of  \ca~4800 m.w.e. It consists of a tracking (wire chambers filled with an ethyl-alcohol mixture, operated in the Geiger mode) and a calorimetric (1940 plastic scintillators) system operated in a 30 gauss magnetic field. A well designed source system allows the simultaneous analysis of up to 10 kg of \BBz active isotopes. Despite a relatively modest energy resolution, implying a non negligible background contribution from \BBd, a sensitivity on \mnu of the order of 0.1 eV has been claimed by the authors on the basis of an excellent control of the backgrounds. As for its previous versions, \BBd remains a primary goal. Data taking will start by fall.
\\
\\
In January 2002, few members of the HM collaboration claimed evidence for \BBz\cite{KDHK0} with $T^{0\nu}_{1/2}=0.8-18.3\times 10^{25}$ y (best value $T^{0\nu}_{1/2}=1.5 \times 10^{25}$ y) corresponding to a \mnu range of 0.11-0.56 eV (best value 0.39 eV). The result steamed from a reanalysis of the HM data based on: i) an automatic peak detection method; ii) identification of the found lines; iii) narrowing of the fit interval to exclude any contribution from recognized line. Due to the importance of such a result, a more extensive substantiation and review (than the one offered in the paper) were asked in a dedicated paper signed by a number of \BB researchers\cite{AAL02}. Moreover, a definitely weaker evidence was found in a next paper repeating the same kind of HM-data reanalysis\cite{VIS02}. In particular, a strong criticism to the validity of the lines identification (inconsistency with other observed more intense lines) was moved in the same paper. Although many of the questions raised by the two criticism papers have been answered in a following paper\cite{KDHK1}, the crucial point of the lines identification is still weak: some recognized line has not been yet identified while the others are still inconsistent or just compatible (in the limit of statistical significance) with expectations. Such a point has been underlined also in a further paper signed by another of the claim authors\cite{KDHK2}.
It is probable therefore that a definite answer to the correctness of the claim will be given only by the very sensitive next generation \BBz projects.

\section{Future Projects}
Most of the criteria that need consideration when optimizing  the design of a new \BBz experiment follow directly from eq.~\ref{eq:sensitivity}: 
\begin{description}
\item[i.] a well performing detector (e.g. good energy resolution and time stability) giving the maximum number of informations (e.g. electron energies and event topology); 
\item[ii.] a reliable and easy to operate detector technology requiring a minimum level of maintenance (long underground running times); 
\item[iii.] a very large (possibly isotopically enriched) mass, of the order of one ton or larger; 
\item[iv.] an effective background suppression strategy.
\end{description}

Unfortunately, these simple criteria are often incompatible and thus no past experiment nor future project could optimize each of them simultaneously. So far, the best results have been pursued exploiting the calorimetric approach which characterizes therefore most of the future proposed projects.
Actually, a series of new proposals has been boosted by the recent renewed interest in \BBz following neutrino oscillation results. I will try to classify them in three broad classes:
\begin{description}
\item[1.] Dedicated experiments using a conventional detector technology with improved background suppression methods (e.g. GENIUS, MAJORANA). 
\item[2.] Experiments using unconventional detector (e.g. CUORE) or background suppression (e.g. EXO) technologies.
\item[3.] Experiments based on suitable modifications of an existing setup aiming at a different search (e.g. CAMEO, GEM)
\end{description}
Expected sensitivities of the proposed projects are compared in Tab. \ref{tab:BBfut}. In some cases technical feasibility tests are requested, but the crucial issue will be the capability of each project to pursue the expected background suppression.
\begin{table}[htb]
\caption{Expected 5 y sensitivities of future projects. NME are from ref. \cite{STA90} except when noted.}
\label{tab:BBfut}
\begin{tabular}{@{}lccc} \\ 
\hline
Experiment                 &Isotope	& $T_{1/2}^{0\nu}$    & $\langle m_{\nu} \rangle$\\ 
                 &	& ($10^{26}$ y)    & (meV)\\ 
\hline
CUORE\cite{AVI01}          &$^{130}$Te  & 7 & 27  \\
CUORICINO\cite{AVI01}     &$^{130}$Te  & 0.15 & 184  \\
EXO\cite{DAN00b}            &$^{136}$Xe  & 8 & 52  \\
GENIUS\cite{KLA01}        &$^{76}$Ge   & 100 & 15   \\
MAJORANA\cite{AAL01}       &$^{76}$Ge   & 40 & 25  \\  
GEM\cite{ZDE01}            &$^{76}$Ge   & 70 & 18    \\
MOON\cite{EJI00}           &$^{100}$Mo  & 10 & 36  \\   
XMASS\cite{MOR01}          &$^{136}$Xe  & 3 & 86   \\
COBRA\cite{ZUB01}          &$^{130}$Te  & 0.01	& 240   \\
DCBA\cite{ISH00}           &$^{150}$Nd  & 0.15	& 190     \\
NEMO 3\cite{SAR00}         &$^{100}$Mo  & 0.04 	& 560   \\
CAMEO\cite{BEL01}          &$^{116}$Cd  & $>$ 1 & 69  \\
CANDLES\cite{KIS01}        &$^{48}$Ca   & 1 &  158\cite{PAN96}   \\
\hline
\end{tabular}
\end{table}

Too many proposal have been recently suggested for a detailed description. We will therefore mention just few selected examples showing the main characteristics of the future \BBz challenge while giving just a very short description of the concept for the others.
\\
Both projects are large scale extensions of existing successfull experiments. 
GENIUS\cite{KLA01} (GErmanium in  liquid NItrogen Underground Setup) will consist of an  array of 400 isotopically enriched (86\% in $^{76}$Ge) Ge diodes, with a total mass of \ca 1 ton. Evolved from the HM experiment, it will aim at a radical background suppression (mainly due to environmental and setup radioactivity) through the use of an unconventional ``cryostat'': naked diodes will be suspended in the centre of a very large liquid nitrogen container, which will act also as a very effective shield.
Liquid nitrogen is in fact available at extremely good radiopurity levels and a reduction by a factor of \ca 100 with respect to the HM background level is expected in the \BBz region.
A problem could be represented by the huge dimensions (12 m diameter) and security requirements  (it would operate underground). Cost and availability of the enriched Germanium are also important issues. The suggested extension to 10 tons, while maintaining the same level of background, would lead to a sensitivity about twice better. 
Three small naked Ge diodes were tested in a small (50 l) LN cryostat for a short time, indicating a performance comparable to that in a conventional vacuum-tight cryostat.
In order to test the feasibility of the project (long time performance of naked Ge diodes in LN and dark matter  studies), the authors have proposed the construction of a preliminary test Facility  (GENIUS-TF).
Already approved by the LNGS Scientific Committee of the Gran Sasso Laboratory, it will consist in 14 crystals of naked natural germanium diodes inside a small liquid nitrogen box. A standard shield (heavy layers of copper, low radioactivity lead and borated polyethylene) will surround it. The GENIUS-TF shield smallness will probably prevent any direct check of the GENIUS background suppression concept, while its large detector mass will probably allow to investigate the Dark Matter annual modulation.
\\
MAJORANA\cite{AAL01}, which involves many of the IGEX collaborators, would also consist of an array of 210 isotopically enriched Ge diodes for a total mass of 0.5 tons. As opposite to the GENIUS design, the use of a very low activity conventional cryostat (extremely radiopure electroformed Cu) able to host simultaneously a number of diodes is proposed. The compact set-up would be installed in the new Underground Laboratory being planned in the USA. 
The driving principle behind the project is a strong reduction of the background by the application of a very effective pulse-shape discrimination and the development of special segmented detectors. The authors believe in fact that main contributions to \BBz background be due to cosmogenically produced long-lived isotopes.
After a number of preliminary tests, the authors are presently mounting a 12 sections segmented enriched crystal underground in order to test their background expectations. 
 They are very concerned by the cost and by the time required for the production and delivery of the required large amount of enriched material. Negotiation are in progress with russian enrichment institutes. The possibility to build a new dedicated enriching facility is under study.
\\
CUORE\cite{AVI01} ({\em Cryogenic Underground Detector for Rare Events}) would be a very large extension of MIBETA also installed in the Gran Sasso Laboratory. CUORE would consist in a rather compact structure made of 1000 cubic natural \teod crystals of 5 cm side (with a mass of 760 g), arranged into 25 separate {\em towers} (10 {\em planes} of 4 crystals each) and operated at a temperature of ~10 mK. The expected energy resolution is \ca5keV FWHM at the \BBz transition energy (2.528 MeV). A background level lower by a factor of 100 with respect to the MIBETA one, is expected by extrapolating MIBETA background results to the CUORE structure. A further improvement is expected on the ground of a better surface contribution suppression. \BBz would be its main investigation but dark matter searches are also foreseen (annual modulation of the signal, axions from the sun, etc.). Thanks to the bolometers versatility, alternative options with respect to \teod could be taken into consideration.
A smaller, but still sizeable, experiment named CUORICINO ({\em small CUORE} in italian), has been already approved and funded and is presently being installed. It consists in a modified single tower of CUORE made by  44 CUORE crystals (11 {\em planes}) plus 18 MIBETA crystals (2 further {\em planes}). 
Even if single CUORE {\em planes} have been already successfully tested (energy resolutions of \ca1 keV at low energy and \ca3-6 keV in the \BBz region over running time of the order of months), CUORICINO will be a crucial test of the CUORE project feasibility (technical performance and background level expectations). 
\\
EXO\cite{DAN00b} would be a large mass (\ca~10 tons) Enriched Xenon Observatory aiming at a $^{136}$Xe \BBz search through an ingenuous tagging of the doubly charged Ba isotope produced in the decay ($^{136}Xe\to^{136}Ba^{++}+ 2e^-$), which would allow an excellent background suppression. The concept of this unconventional proposal is the following: after reduction to Ba$^+$ ion, excitation from the initial $6~^2S_{1/2}$ state to the $6~^2P_{1/2}$ is obtained by means of a first 493 nm laser pulse. Such a state would then decay with a 30\% B.R. to the metastable $5~^4D_{3/2}$ state which can be re-excited to the $6~^2P_{1/2}$ by a second 650 nm laser beam. De-excitation to the original $6~^2P_{1/2}$ state would then be followed by the emission of a 493 nm photon.
The technical feasibility of such an ambitious project aiming at a complete suppression of all the backgrounds requires a hard R\&D phase. The unavoidable \BBd contribution is however a serious concern due to the poor energy resolution of Xe detectors.
Two detector concepts have presently been considered: a high pressure gas TPC and a LXe chamber. In the gas TPC option, Ba ions would remain for a reasonable time in the same position (0.7 mm/sec diffusion at 5 bars), allowing an effective tag (\ca~100 laser cycles) after their position would be measured in the TPC. 
The LXe option would have, on the other hand, the advantages of a more compact structure and of a better energy resolution (scintillation readout) but at the cost of an insufficient spatial resolution. Ions transport into a spectroscopy chamber for a later analysis is under study. 
EXO has been currently funded to develop a 100 kg enriched Xe withouth Ba tagging.
\\
Based on a passive source approach the MOON project\cite{EJI00} plans to use natural molybdenum (9.63~\%) to detect not only \BB but also solar neutrinos. To be installed in the Oto laboratory (Japan), it would consist in a gigantic sandwich (m$_{Mo}$=34 t) made by sheets of natural molybdenum interleaved with specially designed scintillators. The possibility to use bolometeric detectors has also been considered. The CAMEO\cite{BEL01} proposal would use 1 ton of scintillating $^{116}$CdWO$_4$ cristals inside the Borexino detector. CANDLES would be based instead on the use of CaF$_2$ in liquid scintillator. COBRA\cite{ZUB01} would use CdTe or CdZnTe diodes (\ca10 kg) to investigate Cd and Te \BB isotopes in a calorimetric approach. GEM\cite{ZDE01} is a proposal very similar to GENIUS in which the complex LN huge cryostat has been replaced by a definitely smaller one inserted in a large pure water container (e.g. Borexino). DCBA\cite{KIS01} proposes the use of a modular 3-D tracking (drift chamber) in a uniform magnetic field to study $^{150}$Nd \BBz; the expected sensitivity based on the  analysis of the single electron energy distributions seems unfortunately untenable. An interesting \BBz sensitivity have been claimed also by the XMASS\cite{MOR01} solar neutrino collaboraion.

\section{Conclusions}
A renewed interest in \BB has been stimulated by recent neutrino oscillation results. Neutrinoless \BB is finally recognized as a unique tool to measure neutrino properties (nature, mass scale, intrinsic phases) unavailable to the successful experiments on neutrino oscillations. 
Present limits on \mnu are still outside the range predicted on the basis of the latest neutrino oscillation results. However the situation could drastically change in the future. A number of newly proposed experiments could in fact reach this sensitivity.
The attainability of such a goal strongly depends on the true capability of these projects to reach the required background levels in the \BBz region. An experimental confirmation of the (sometimes optimistic) background predictions of the various projects (even if extrapolated from the results of lower scale successful experiments) is therefore worthwhile and the construction of preliminary test setups is absolutely needed. These could be on the other hand experiments at an intermediate scale (e.g. GENIUS-TF or CUORICINO).
The recently claimed evidence for a \BBz signal in the HM data seems still too weak but could be verified in the next future experiments.\\
A strong effort to improve the NME evaluation should be encouraged while stressing on the need of experiments addressed to different nuclei.
The possibility to exploit \BB decay experiments to investigate different processes (as often observed in the past) should also be stressed.

\end{document}